# Conversational Forecasting Across Large Human Groups Using a Swarm of Surrogate AI Agents

Louis Rosenberg
Unanimous AI
Pismo Beach, CA
ORCID: 0000-0003-3457-1429

Hans Schumann
Unanimous AI
San Francisco, CA
Hans@Unanimous.ai

Ganesh Mani
Carnegie Mellon University
Pittsburgh, PA
ORCID: 0000-0002-2170-7414

Gregg Willcox
Unanimous AI
Seattle, WA
Gregg@Unanimous.ai

*Abstract*—**Hyperchat AI is a communication and collaboration architecture that employs intervening AI agents to enable real-time conversational deliberations among networked human teams of unlimited size. Prior work has shown that teams as large as 250 people can hold productive real-time conversations by text, voice, or video using Hyperchat AI to discuss complex problems, brainstorm solutions, surface risks, assess alternatives, prioritize options, and converge on optimized results. Building on this prior work, this new study tasked groups of 25 to 30 basketball fans with conversationally forecasting NBA games (against the spread) over a 12-week period. Results show that when discussing and debating NBA games (for five minutes each) using a Hyperchat AI enabled platform called Thinkscape, human teams were 62% accurate across a set of 50 forecasted NBA games. This is an impressive result versus the Vegas odds of 50% (p=0.059). Furthermore, had the participants wagered on the games, they would have produced an 18.4% ROI over the 12-week period. In addition, this study found that the group's conversation rate during each forecast was positively correlated with their prediction accuracy. In fact, when excluding the 12 forecasts in the bottom 25th percentile by average conversation rate, the remaining 38 forecasts recorded a 68% accuracy, significantly better than the 50% Vegas odds (p=0.017). This result also outperformed the well-known prediction market Polymarket (p=0.062) across the same set of NBA games. These outcomes suggest that real-time conversational deliberations, when facilitated by Surrogate AI agents, can significantly amplify groupwise collective intelligence during human forecasting tasks.**

## I. INTRODUCTION

From sales and marketing, to engineering and operations, teams within large organizations often need to generate accurate forecasts that leverage the diverse knowledge, expertise, insight, and situational awareness of distributed team members. In many organizations today, this is achieved by collecting data from individual stakeholders through polls, surveys, or interviews (a process often called "expert elicitation"). While useful, these methods treat each participant as a source of isolated input for aggregation, providing no mechanism for deliberative reasoning among members and no pathway for participants to collectively refine their thinking by discussing and debating their views, knowledge, insights, and rationales with each other [1,2].

Hyperchat AI is a novel agentic architecture that addresses this need by enabling large networked teams (of any size) to hold productive real-time deliberations by text, voice, or video. The goal is to enable stakeholders within large organizations to efficiently discuss and debate issues and quickly converge on accurate insights, decisions, estimations, assessments, forecasts, prioritizations, or solutions [3]. Hyperchat AI™ technology (also called Conversational Swarm Intelligence) achieves this by dividing a large networked team into a set of small deliberative subgroups that are linked together by a swarm of innovative AI agents called "conversational surrogates." These agents track, process, and connect the parallel discussions in each subgroup, enabling large teams to hold thoughtful real-time conversations in which they brainstorm together, prioritize together, debate and assess risks, and ultimately forecast outcomes together that benefit from their diverse knowledge, skills, and insights [4].

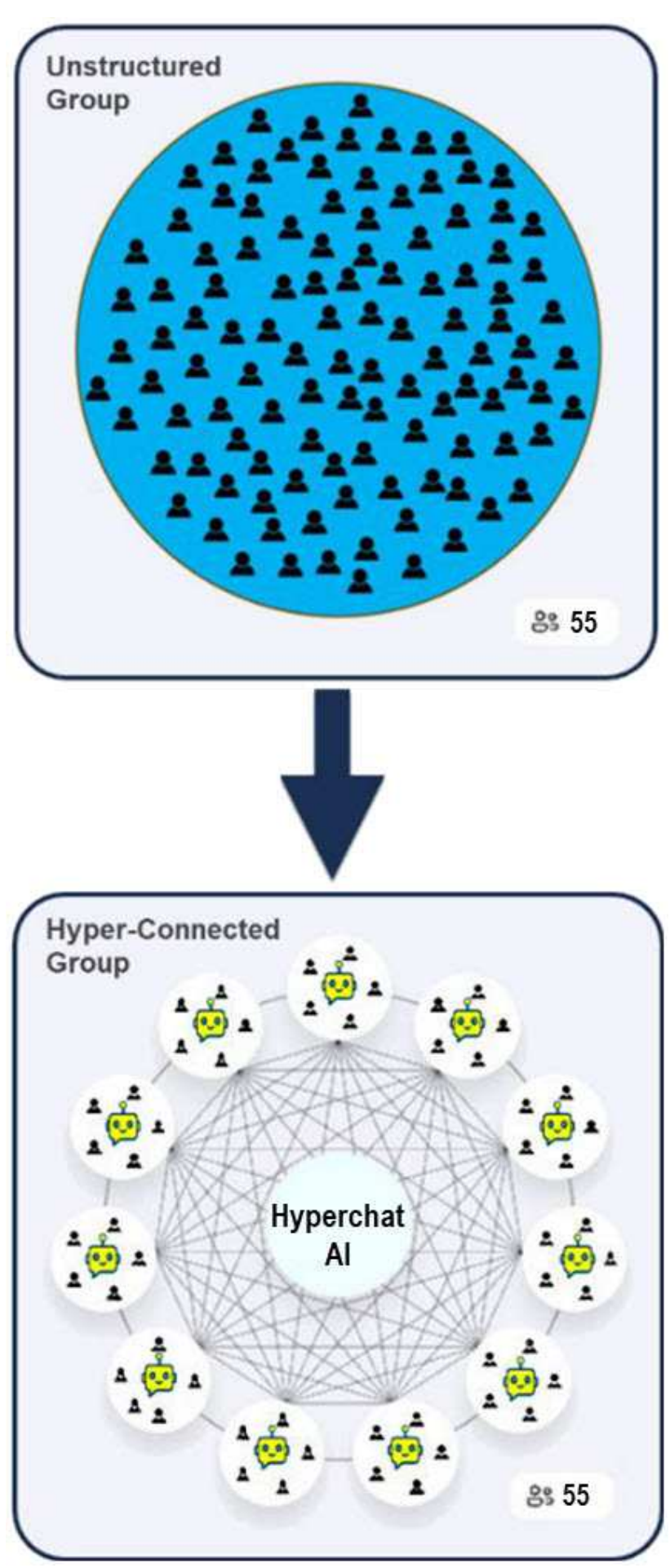

Fig 1. Hyperchat AI enables thoughtful deliberations at scale

## II. FROM SWARM AI TO HYPERCHAT AI

Hyperchat AI is modeled on the biological principles of Swarm Intelligence. This is the natural phenomenon that enables large groups of social organisms (such as bird flocks, bee swarms, and fish schools) to quickly and efficiently reach accurate collective decisions. Unlike traditional human methods for harnessing collective intelligence, biological swarms do not simply collect and aggregate individual perspectives but instead enable groups to form a real-time control system that pushes and pulls on the decision-space and converges in unison on solutions that maximize their collective support [5].

The first methodology to enable networked human groups to form a swarm intelligence was developed in 2014 and is called Artificial Swarm Intelligence (ASI) or simply Swarm AI® [6]. Using this technology, researchers showed that human teams could form interactive dynamic systems modeled on biological swarms and when collaborating this way, significantly increased groupwise accuracy when performing tasks such as predicting financial markets, estimating product sales, generating medical diagnosis, and making sports predictions [7,8,9].

Although teams using Swarm AI technology have shown significant intelligence amplification, one drawback is that the deliberations are graphical, not conversational. Because of this, ASI-based Swarm® collaboration platforms enable large groups to quickly converge on optimized solutions but do not reveal the underlying reasons and rationales that drive participants to the solutions they converge upon. This was solved in 2023 with the creation of Hyperchat AI (also referred to as Conversational Swarm Intelligence or CSI) which enables the intelligence amplification benefits of real-time biological swarms but does so through natural real-time conversations among large human groups mediated by intervening AI agents [3,10, 13].

In 2024, a decision-making study was conducted using Hyperchat AI in which groups of 25 to 30 participants were asked to select players for a weekly Fantasy Football contest [11]. Researchers compared player selections using a traditional aggregated survey to player selections made using real-time "conversational swarms." Results showed that the groups using Hyperchat AI outperformed 66% of survey participants, revealing intelligence amplification versus the median individual ($p=0.020$). The groups using Hyperchat AI also significantly outperformed the aggregated survey (i.e., the most popular choices) ($p<0.001$). These early results suggest that Hyperchat AI is an effective architecture for group decision-making, but more rigorous forecasting studies were deemed necessary to adequately validate this new methodology.

In addition, to support rigorous conversational forecasting, the Hyperchat AI engine needed to be extended with additional capabilities that enable AI agents to track the unique forecasted values expressed by users in each of the parallel subgroups along with the strength of their conversational sentiment, and identify the unique points and counterpoints raised by participants as to why the proposed values may be too high or too low. This real-time tracking and processing enable the Hyperchat AI to build a dynamic database of forecasted values linked to reasons and rationales that justify each value. The real-time database is then used to optimize information sharing between subgroups, as conducted by the conversational surrogate agents.

In addition, the Hyperchat AI engine tracks changing sentiments across the full population of participants and shares this using a real-time graphical display. This allows participants to quickly assess the collective views of other participants as they conversationally discuss and debate the optimal forecast value. Specifically, the Hyperchat AI engine tracks and displays collective sentiments at local, regional, and global levels within the hyper-connected structure. When displaying sentiment profiles at the *local level*, participants see a graphical depiction of the collective forecasting sentiments across the participants in their own subgroup. When displaying sentiment profiles at a regional level, the participants see a graphical representation of the forecasting sentiments across their own subgroup and one or two other subgroups. And when displaying sentiment profiles at the global level, participants see a graphical representation of the forecasting sentiments across the full population of participants. The steadily expanding level of information sharing mitigates social influence bias by limiting the scope during early dialog but gradually expanding the scope as the conversation evolves, with more and more deliberative connections being made between subgroups by intervening AI agents.

## III. NBA FORECASTING STUDY

The goal of this study was to quantify the effectiveness of collaborative group forecasting among networked human teams via conversational deliberations facilitated by Hyperchat AI technology. The study used the Thinkscape.ai software platform from Unanimous AI configured to enable real-time text-chat deliberations among networked teams of randomly selected sports fans. In total, 56 NBA games were collaboratively predicted during a twelve-week period of the 2025-2026 NBA season. The groups of sports fans were sourced from a professional panel-providing service (Prolific). Each participant self-identified as a "basketball fan" but was otherwise randomly selected. Each was paid approximately $6.50 for taking part in a single forecasting session in which four games (scheduled to be played that same day) were predicted against the spread. The collaborative forecasts were conducted conversationally, with each group being allocated approximately 5 minutes for real-time text-based debate and discussion regarding each game.

Each group of 25 to 30 participants was shown a 30-second instructional video about how the Hyperchat AI architecture works. Specifically, they were told they would be randomly divided into five or six subgroups (called Thinktanks), each with approximately 4 or 5 participants. As described above, they were informed that each Thinktank enables real-time conversation among the 4 or 5 human members and a single AI agent called a Conversational Surrogate. Each AI agent was tasked with continuously (a) <u>monitoring</u> the local conversation within its subgroup, (b) <u>identifying</u> the key insights raised by participants within its subgroup conversation, (c) <u>sending</u> key insights to the Surrogate Agents in other subgroups, and (d) when receiving insights from agents in other subgroups, <u>expressing</u> the received insights as natural collaborative dialog within its own subgroup. It should be noted that each "insight" that is identified, shared, and expressed in this way, consists of a forecasted <u>outcome</u>, one or more <u>reasons</u> that support or oppose that forecasted outcome, and an inferred level of <u>confidence</u> or <u>conviction</u> based on the interactive dynamics of the conversational deliberation.

It is also important to note that the core steps of *identifying* and *sharing key insights* are managed by an oversight AI process called the "Deliberative Matching Engine" or DME. This subsystem is designed to select and share conversational insights among subgroups that are likely to have maximal impact on the receiving group. This is achieved by tracking every unique insight that each subgroup has been exposed to thus far in the deliberation, either because it was raised organically within that subgroup or because it was shared into the subgroup by an AI agent. In this way, the DME engine tracks real-time "exposure" of subgroups and only shares conversational content that is novel to each subgroup. This optimizes information passing.

In addition, the DME engine tracks the prevailing opinion within each subgroup based on the real-time dialog among the subgroup members. When selecting a novel insight to be shared with a given subgroup, the selected insight is chosen to maximally challenge the receiving subgroup. In this way, each subgroup is exposed to ideas, insights, reasons, and rationales that are most likely to impact their ongoing conversation, either by (i) swaying their opinions in an alternate direction, or (ii) by inspiring members of the receiving subgroup to push back on the received insight expressed by their surrogate agent. In both cases, this evokes useful information to the oversight system as it reveals which insights are the most influential as they spread around the hyper-structure and which insights are viewed with skepticism as they propagate.

In this way, the "conversational matching" process optimizes the mixing of ideas, insights, reasons, and rationales and inspires groups to push back on prevailing opinions that are emerging in other subgroups that they disagree with by expressing doubts or counterpoints they believe are significant. In turn, these doubts and counterpoints get shared around the network, speeding deliberation. This enables a large group of dozens or even hundreds of collaborators to quickly explore an issue, raising supporting and opposing views, all while AI agents ensure that the full population is efficiently exposed.

Using these methods, the swarm of conversational surrogate agents connect the set of parallel local discussions into a single global conversation in which diverse perspectives are efficiently debated, for or against each team, with information and arguments being continually shared across subgroups. For example, if the members of a subgroup collectively believe that the Minnesota Timberwolves will beat the Portland Trailblazers by more than the 5 point spread published by oddsmakers, the AI agent in that subgroup will likely receive and express counterpoints sourced from other subgroups that have not yet been considered locally. This process occurs simultaneously in all subgroups, ensuring that information and insights, both supporting and opposing each outcome, efficiently propagate as natural dialog across the full collaborative group.

Using these methods, a set of four NBA games were forecast conversationally during each test session using the current spreads published by DraftKings. The participating group of 25 to 30 sports fans were automatically divided into five or six subgroups and were allocated approximately 5 minutes to deliberate each game. Therefore, each forecasting session lasted approximately 20 minutes and resulted in four predictions, each generated entirely via real-time conversational deliberations conducted as text chat. That said, the Hyperchat AI engine enables the same process by teleconference or videoconference. The reason text chat was used in this study is that it preserves anonymity for paid participants.

Each collaborative forecast begins with a text prompt provided by a human moderator that indicates the question to be answered and any relevant guidance. In this study, each question identified a pair of NBA teams playing that night along with the current spread (Figure 2 shows a screenshot from the study).

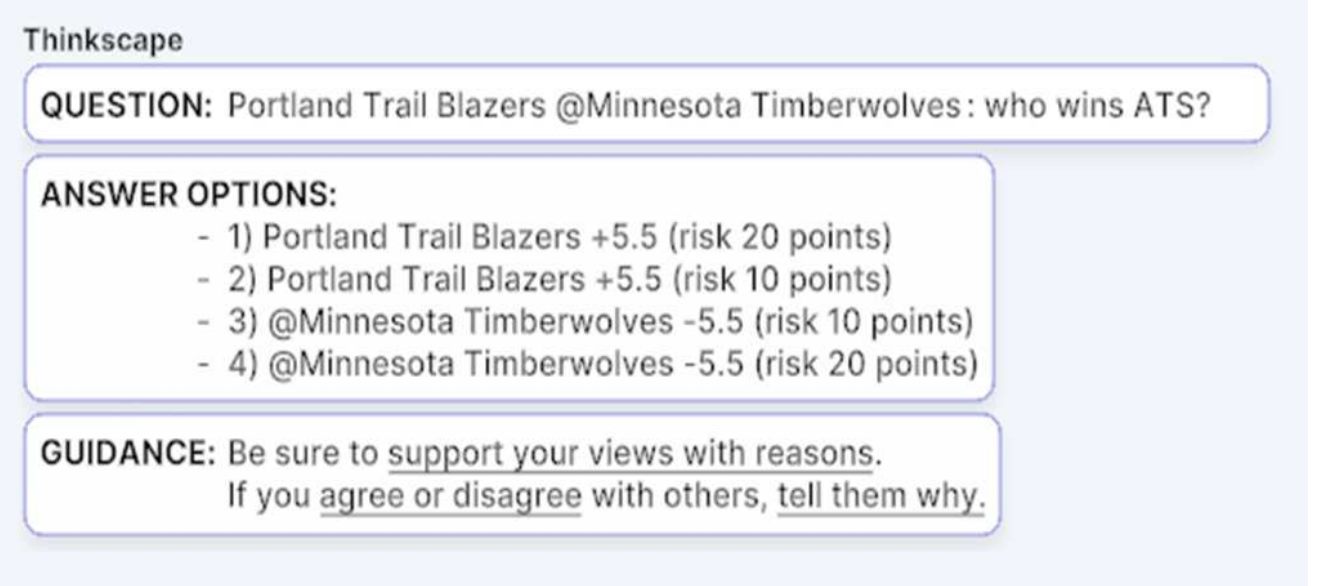

Fig 2. Screenshot of forecasting question presented to participants

As shown in Figure 2 above, the forecasting question was structured to elicit thoughtful groupwise conversation regarding which team was most likely to beat the spread and why. In fact, the guidance indicates very clearly that participants should support their views with reasons and if they agree or disagree with others, they should express those sentiments. The reason is to ensure the conversation is an interactive deliberation, with arguments and counter-arguments, not just a set of opinions.

As also shown in Figure 2, each forecasting question was structured with four options the group must choose among, each of which is associated with one team beating the spread and a level of risk (10 points or 20 points) that should be "collectively wagered" on that team. The reason for having two wager levels is to differentiate between "high confidence" and "low confidence" forecasts. This encourages thoughtful conversation, even in situations where a large majority of participants believe that one of the teams is likely to beat the spread. In such situations, the group will deliberate over whether they should risk 10 or 20 points on this outcome. Prior research has confirmed that this structure yields deeper deliberations that allow the Hyperchat AI engine to more accurately quantify the perspective and confidence of each individual, each subgroup, and the full population of participants [4, 12].

At the start of each forecast, a timer appears on each user's screen that counts down from 5 minutes, thereby indicating how much time is left to discuss the game. As the set of parallel deliberations unfold, each participant's dialogue is processed by a Large Language Model (LLM) to determine (i) the team they believe is most likely to beat the spread, (ii) the strength of their conviction in that belief, and (iii) their reasoning for holding that belief. This occurs in parallel for all 25 to 30 participants and is continuously updated as the discussion transpires. This allows the Hyperchat AI engine to track how each individual forecaster shifts their views and/or adjusts the strength of their conviction as the conversation unfolds. The Hyperchat AI engine tracks these changing sentiments for individuals, subgroups, and the full population at all moments in time. Figure 3 below shows a plot of aggregated sentiment across the full population over the five-minute deliberation for a forecasted game (i.e., the Timberwolves vs Trail Blazers at a 5.5 spread):

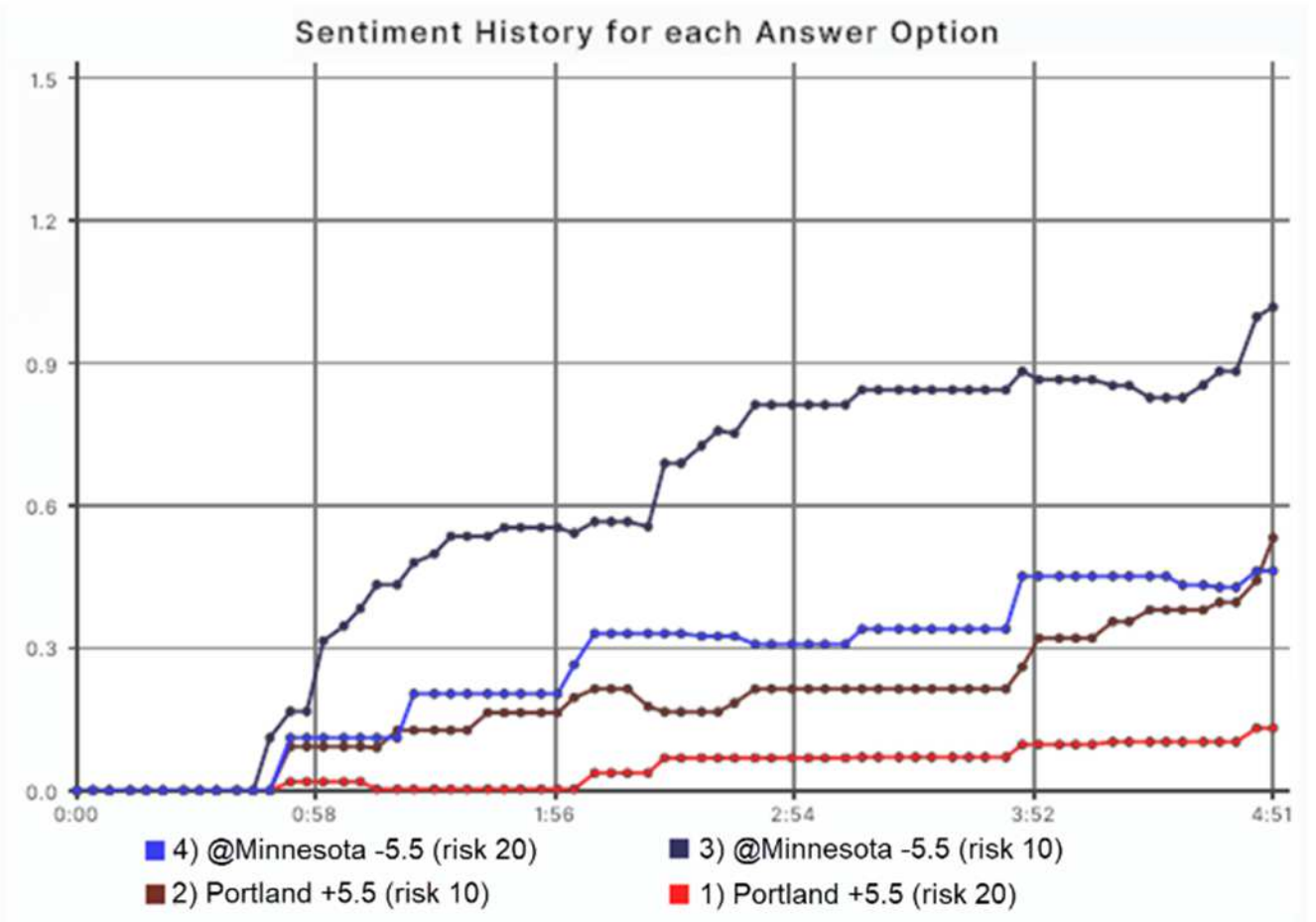

Fig. 3. Screenshot of real-time conversational sentiment tracking in Thinkscape

In addition, the Hyperchat AI engine tracks (in real-time) the conversational support for each of the four choices that the group is debating. It then uses this information to provide participants with an up-to-date graphical representation of support across the population of participants (see Figure 4 below). In addition, the Hyperchat AI engine computes a weighted mean across this support profile and plots the mean as a dotted line on the graph. This gives participants a real-time indication of the group's changing perspectives based on the ongoing deliberation.

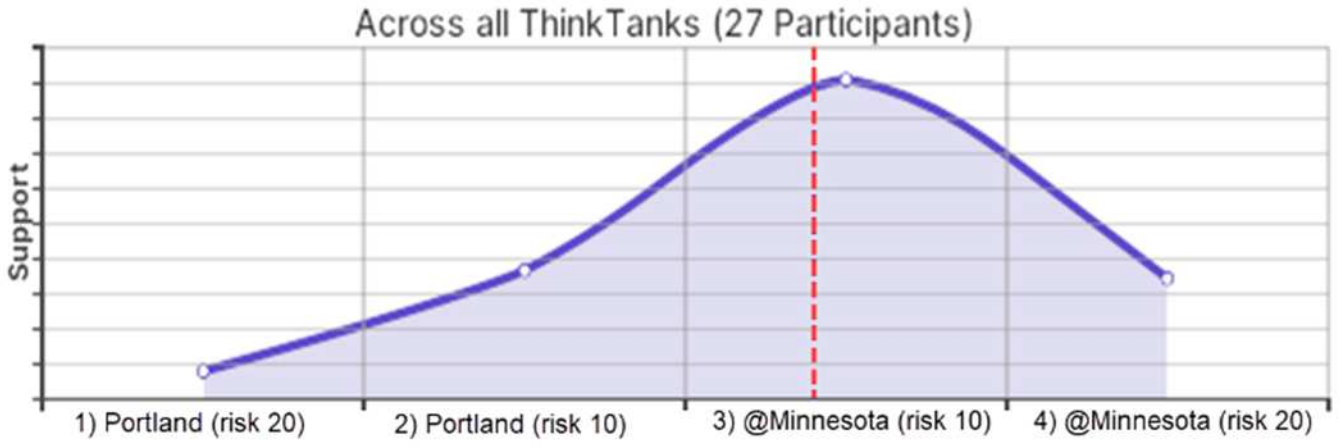

Fig. 4. Support Profile Across All Thinktanks (27 Participants) in real-time.

The ability to provide a real-time graphical representation of conversational sentiment (with a computed mean) gives users the ability to comment on the current collective forecast, for example expressing to their subgroup whether they believe it is too high or too low and why that to be the case.

## IV. Data, Analysis, and Results

During each deliberation, each user's conversational dialog is assessed in real-time to determine which team they believe is most likely to beat the spread, Team A or Team B, and which level of risk they believe the group should collectively wager on that team, 10 points or 20 points. Thus, for each user at every moment in time, the system assesses (based on their deliberative dialog) which of the four forecast options they currently support the most: Team A (risk 10 points), Team A (risk 20 points), Team B (risk 10 points), or Team B (risk 20 points), and assigns a "Support Value" to that option on a scale of 0 to 100%, with the sum of all Supports totaling 100%. This support value indicates relative conviction in that forecast based on the user's conversational deliberation.

For example, a user who expresses more certainty in their choice for Team A or Team B will be assigned a higher Support Value for risking 20 points than a user who does not express confidence and does not provide compelling reasoning as to why they are confident. It is important to note that these Support Values are based on how the user behaves in the deliberation, which includes not just the declarative statements they make about which team they think will beat the spread, but also how they react when challenged by counterpoints or conflicting evidence, and how their views shift over time. Thus, at the end of each forecasting deliberation, the system will have assessed each user in each separate subgroup and assigned Support Values for that user with respect to each forecast option.

At the conclusion of each conversational deliberation, a Weighted Collective Forecast (WCF) is computed across the full population of participants. The WCF is generated by computing a weighted mean of the Support Values assigned to each of the forecast options, across the full set of participants . This value is computed on a scale of -2 to +2 in which (-2) indicates 100% support for Team A, (+2) indicates 100% support for Team B, and 0.00 indicates a split decision with 50% support for each. As shown in Equation 1 below, the supports (w) for each answer are multiplied by their associated answer value and are summed to get the Weighted Collective Forecast.

$$\bar{x}_w = (w_1 \cdot -2) + (w_2 \cdot -1) + (w_3 \cdot 1) + (w_4 \cdot 2)$$

Eq. 1. Weighted Collective Forecast equation

The Weighted Collective Forecast is used to determine which team the group collectively believes is most likely to beat the spread, A or B (and which risk option, 10 points or 20 points is collectively preferred). If the weighted mean is directly in the center between -2 and +2 (i.e. 0.0) or within a small margin of error around the center, the forecast is classified as a toss-up. In this study, "toss-ups" were defined as Weighted Collective Forecasts that fell between -0.08 and +0.08, on the defined scale of -2 to +2. These toss-ups were rare and were classified as "no forecast" (i.e. the group did not collectively identify one team that was more likely than the other to beat the Vegas spread outside this small margin of noise). Across the 56 forecasts made during this study, only 6 forecasts classified as toss ups, leaving **50 total forecasts** made by the group that could be scored.

For all other picks, the side that the weighted mean falls on determines the pick (i.e., Team A is picked if the mean is less than -0.08, and Team B is picked, if the value is greater than +0.08) with the strength of the pick represented by its proximity to -2 or +2 respectively. As shown in Table 1 below, of the **50** forecasts that identified a preferred team to beat the spread, the collaborative groups got **31** correct (**62%** accuracy). This value is substantially higher than the 50% expected accuracy at the 10% level using a Binomial Test, showing that Thinkscape forms an intelligent system able to outperform Vegas odds.

These results were compared against the individual accuracy of participants. Each participant provided a personal prediction prior to each group deliberation. Across the 12 week study, there was a total of 43 human subjects who each participated in at least 12 game forecasts against the spread. The cumulative accuracy for each of the 43 human subjects was computed and plotted as a distribution in Figure 5 below. In addition, the mean accuracy across the 43 participants was computed and found to be 52%. As mentioned above, the groups forecasting using Hyperchat AI achieved a mean accuracy of 62%, which outperformed 36 of the 43 participants (84%) as plotted in Figure 5 below.

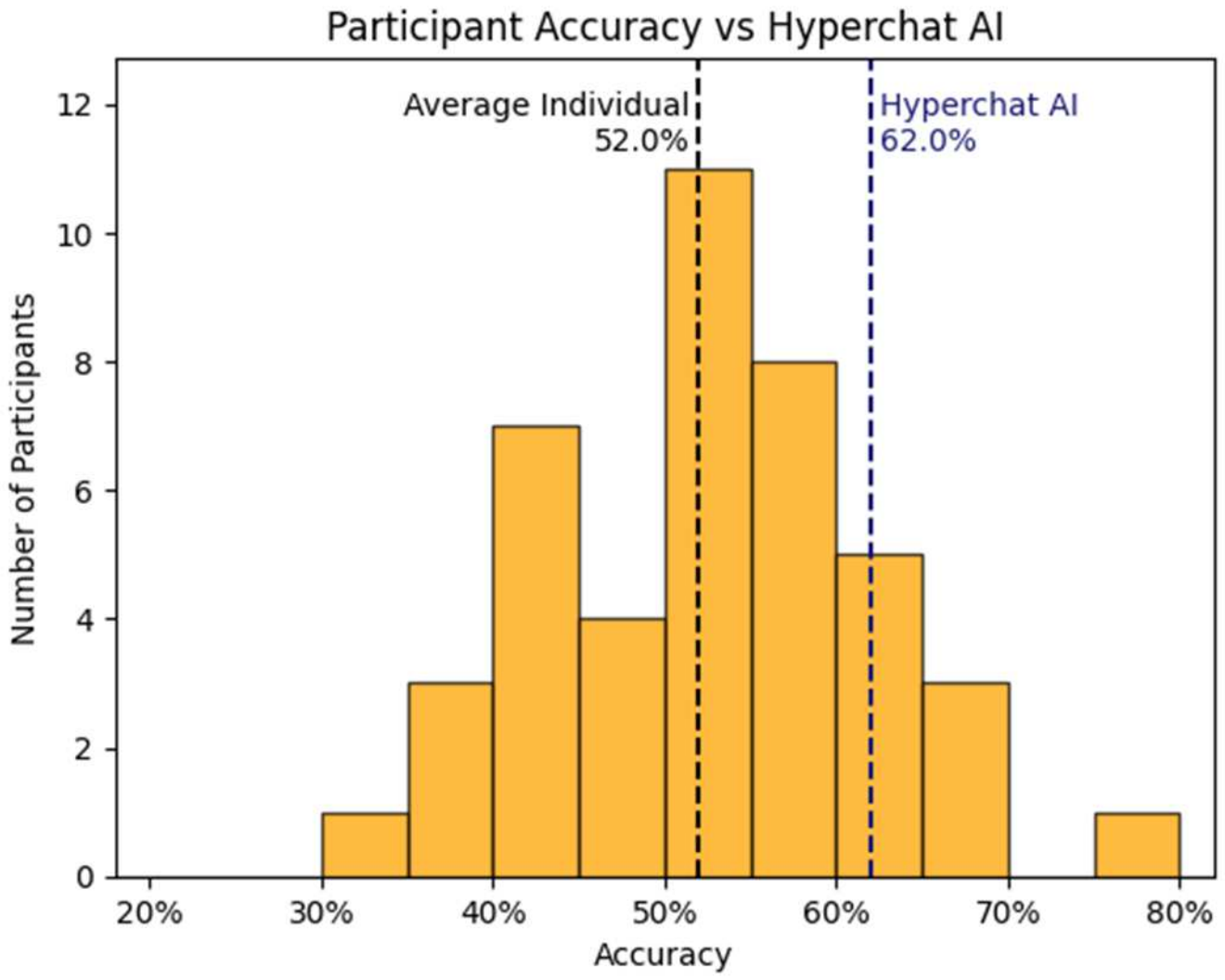

Fig. 5. Accuracy of 43 individual participants vs Hyperchat AI groups.

To further assess the of this accuracy amplification, the set of individual human forecasts were bootstrapped 10,000 times. The bootstrapped distribution and average accuracy is shown in Figure 6. We find that Hyperchat AI forecasts were notably more accurate than the individual forecasts (**p=0.071**).

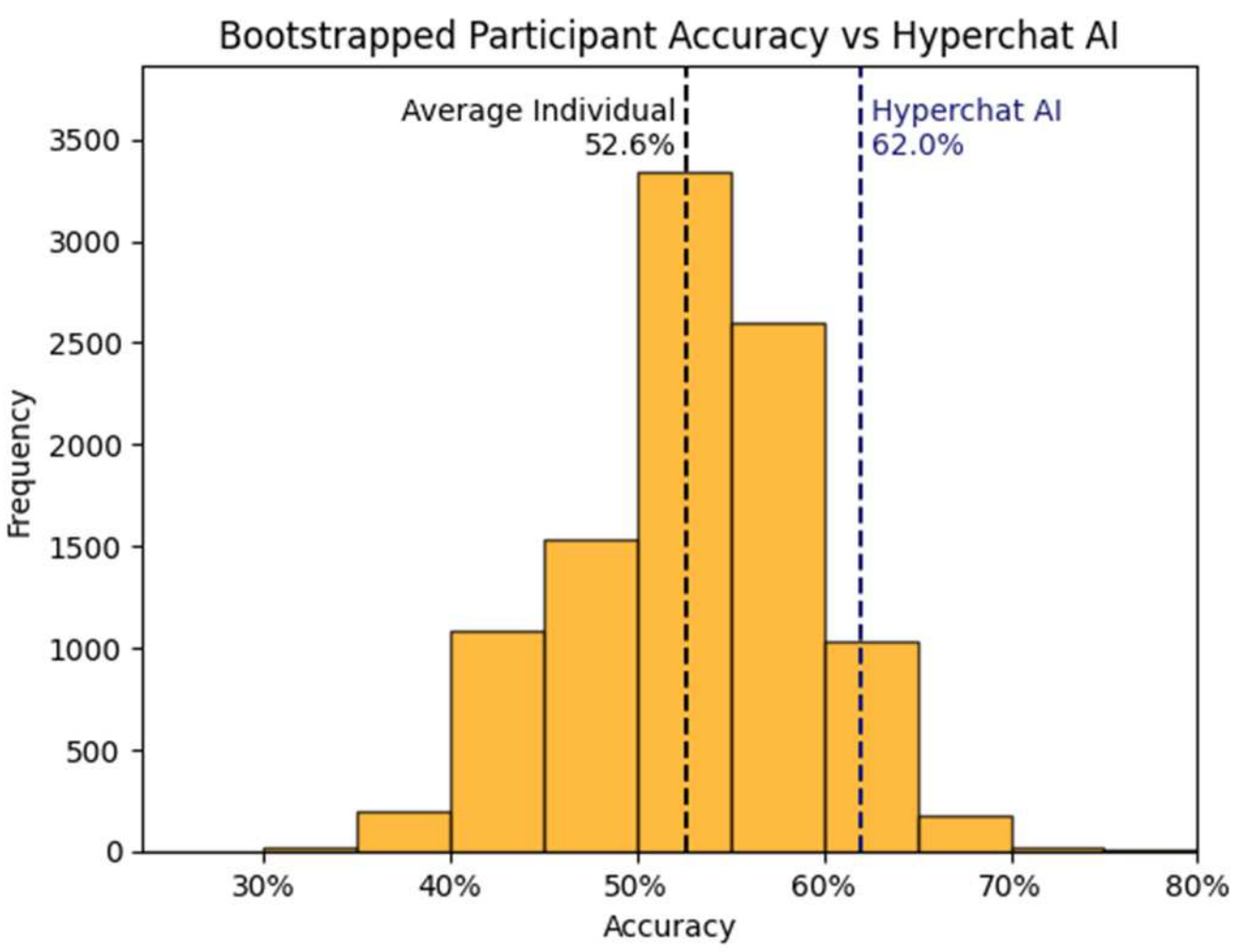

Fig. 6. Bootstrapped Individual Accuracy vs Hyperchat AI groups.

Looking deeper, we find that the groups collaborating using Hyperchat AI outperformed Vegas oddsmakers on both favorite picks (59% accuracy) and underdog picks (73% accuracy) as shown in Table I. In addition, we looked at implied ROI. Each game forecasted had -110 odds on Vegas markets, meaning that a $100 bet on each game would yield $90.91 of profit if correct, and a $100 loss if incorrect. Using these standard payouts, the ROI across all 50 picks was **+18.4%.** In other words, betting $100 on each game using the Hyperchat AI picks would have resulted in $918 of profit over the 12-week study.

TABLE I. ACCURACY OF SWARM BETS

| *Bet Type* | *Num Games* | *Record (Accuracy)* | *ROI* | *p-value* |
|---|---|---|---|---|
| All Picks | 50 | 31-19 (**62.0%**) | +18.4% | 0.059 |
| Favorite Picks | 39 | 23-16 (**59.0%**) | +12.6% | 0.168 |
| Underdog Picks | 11 | 8-3 (**72.7%**) | +38.8% | 0.113 |

Hyperchat AI is designed to amplify intelligence by allowing groups to share information more effectively. While the group was very accurate on all picks made, bad picks could be eliminated by removing forecasts in which the group collectively had less to say. The characters sent per minute per participant is tracked for each forecast, and ranges from 37.9 to 55.6 characters. The 25th percentile, or first quartile, of conversation rate is 43.0 characters per minute per participant.

Using the 25th percentile as a cutoff to signify low conversation rate, we find that pick accuracy differs above and below that cutoff. As shown in Table II below, low conversation rates correlate to weak forecast accuracy, while the greater conversation rates lead to higher accuracy. The upper 75% of conversation rates result in 68.4% accuracy on 38 forecasts, with a significant p-value (p=0.017).

TABLE II. ACCURACY BASED ON CONVERSATION RATE

| *Conversation Rate* | *Num Games* | *Record (Accuracy)* | *ROI* | *p-value* |
|---|---|---|---|---|
| All Picks | 50 | 31-19 (**62.0%**) | +18.4% | 0.059 |
| Lower 25% | 12 | 5-7 (**41.7%**) | -20.5% | 0.806 |
| Upper 75% | 38 | 26-12 (**68.4%**) | +30.6% | 0.017 |

Focusing on the upper 75% of trials by conversation rate, we find (as shown in Figure 7) that the 68.4% forecasting accuracy produced when using Hyperchat AI outperformed **95%** of the participants who took part in the study (by individual accuracy).

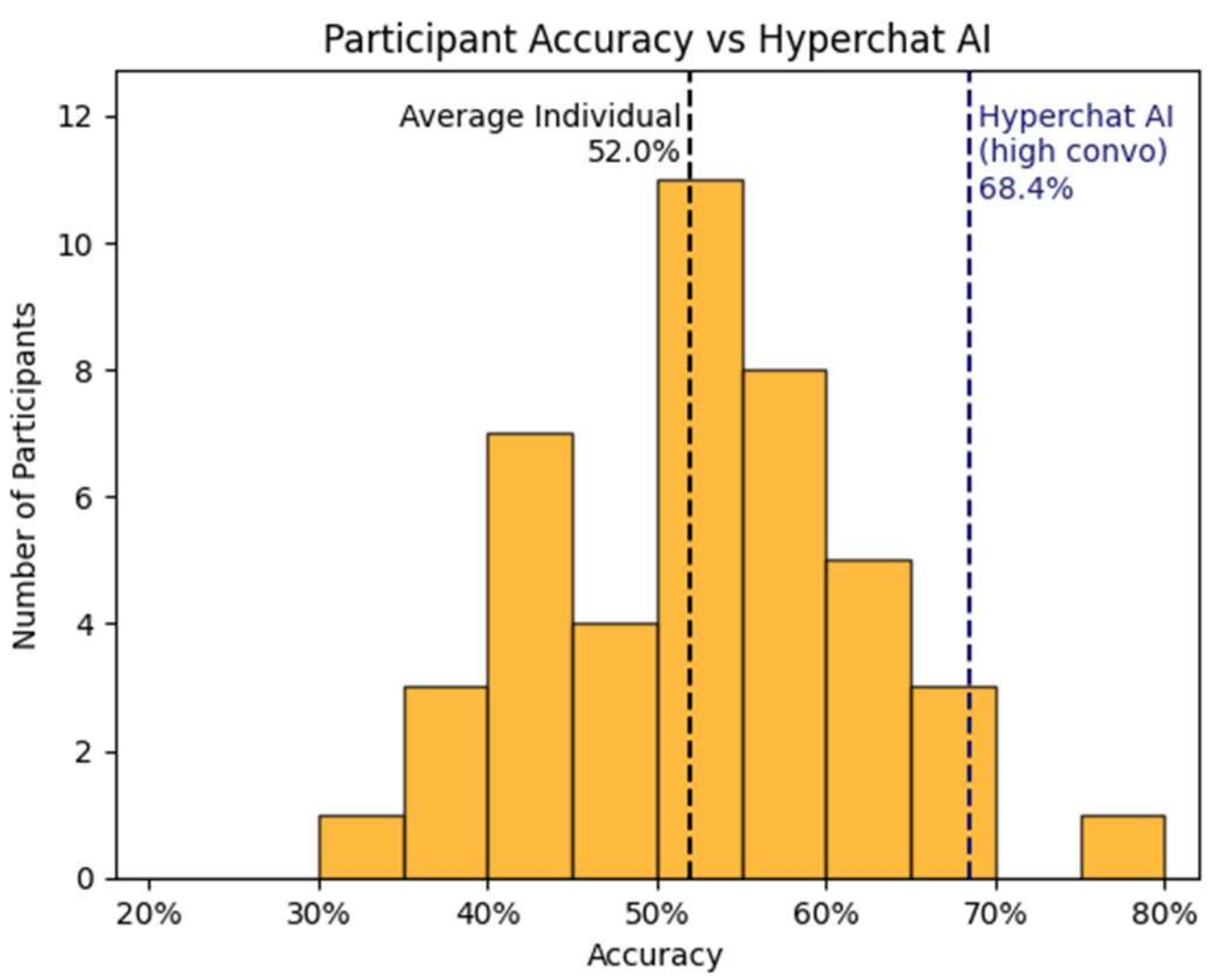

Fig. 7. Hyperchat Accuracy when removing low-conversation trials.

Thus, the collaborating teams using Hyperchat AI were able to forecast NBA games with impressive **62%** accuracy against the Vegas spread. In addition, when removing the low conversation rate forecasts, the teams using Hyperchat AI achieved **68%** accuracy against the spread, which was a significant result vs the Vegas odds (p=0.017). This implies that conversational deliberation improved the collective intelligence harnessed during the groupwise forecasts using Hyperchat AI.

And finally, because of the popularity of prediction markets, we compared the accuracy of the conversational groups using Hyperchat AI (each group with 25 to 30 participants) with a large-scale prediction market. Specifically, we accessed the Polymarket platform using their API and computed the results for the same set of 50 NBA games (against the same spreads). We do not know how many participants contributed to each Polymarket forecast but we can report that average contract value on Polymarket for individual NBA games is currently around $400K to $500K and the average contract size is currently $95 [16]. This suggests that approximately 4,000 to 5,000 unique users participated in each Polymarket forecast.

Across these games, the Polymarket forecasters picked the correct team with **54.8%** accuracy. In other words, more contracts were placed on the team that beat the spread in 54.8% of the predicted games. Thus, we find that the **68.4%** accuracy generated by teams using Hyperchat AI (when deliberating with higher conversation rates) substantially outperformed the 54.8% accuracy generated by Polymarket (p= 0.062) for the same set of games. This is an impressive result, especially considering that each Hyperchat AI forecast only engaged 25 to 30 people, compared the thousands participating in Polymarket forecasts.

Additionally, we compared Hyperchat AI's accuracy to the Polymarket forecasts using a bootstrap analysis of cumulative profit (had contracts been purchased on Polymarket.) By buying 100 shares for each game on the side picked by Hyperchat AI, roughly costing $49 to $51 per game, the resulting profit after the 50 games would be $634. This outperformed 96% of all bootstraps representing a naïve average forecaster, where a random side was chosen for each game (p=0.039).

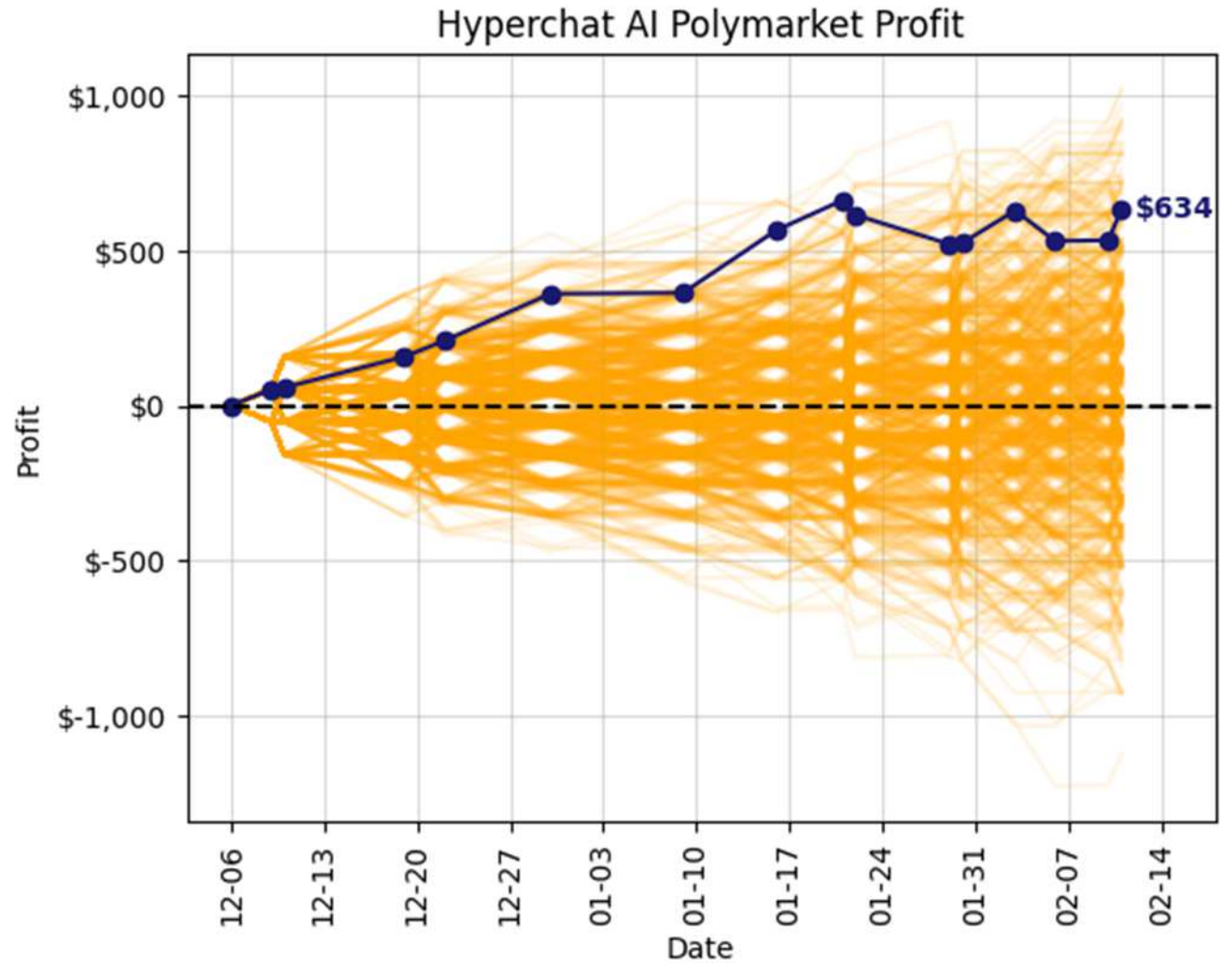

Fig. 8. Hyperchat Polymarket profit compared to bootstrapped average bettors.

## V. Conclusions

This research provides empirical evidence that large-scale conversational deliberations, when mediated by intervening AI agents enabled by Hyperchat AI technology, can significantly enhance group forecasting accuracy in a real-world prediction task. Across 50 NBA games forecast over a 12-week period, groups of 25–30 randomly selected basketball fans achieved **62.0%** accuracy against the spread, outperforming the 50% benchmark implied by Vegas odds (**p = 0.059**). Under standard -110 wagering conditions, this performance corresponds across the 50 predicted came to achieve a **+18.4%** ROI against Vegas, demonstrating practical as well as statistical relevance.

As context for these results, **62%** accuracy against the Vegas spread in sports forecasting is extremely rare to achieve over a large set of games. Industry experts generally indicate that professional sports bettors rarely sustain a long-term win percentage higher than ~55 % and often note that 53–54 % is uncommon against typical spreads with -110 odds [14, 15].

Looking beyond the raw accuracy, this study revealed that conversational forecasting performance (when groups used the Hyperchat AI powered Thinkscape.ai platform) was positively correlated with the average rate-per-minute of conversational deliberation. When the rate of conversation was above the 25th percentile threshold, accuracy rose to **68.4%** (p = 0.017) with a +30.6% ROI (across 38 games). In contrast, forecasts generated during comparatively low conversational engagement (below the 25th percentile), the resulting forecasts had a significantly lower accuracy (41.7%) across those 12 games.

This finding suggests it's not merely aggregation of opinions that drives improved group performance using Hyperchat AI, but the depth and dynamics of large-scale deliberative exchange. In other words, accuracy gains appear to arise from interactive discussion and debate, with participants considering arguments and counterarguments made by other participants, whether expressed directly by that participant or by the intervening AI agents enabled by the Hyperchat AI architecture.

Also, the high conversation rate forecasts were compared to a large-scale prediction market (Polymarket) which aggregates input from forecasters without any deliberative conversational interactions. When predicting the same set of NBA games against the spread, Hyperchat AI method had 68.4% accuracy, which outperformed the computed accuracy on Polymarket 54.8% (p=0.062). This is an exciting result when considering that the Hyperchat AI forecasts engaged only 25 to 30 human forecasters versus thousands for each Polymarket prediction.

These results extend prior work in Artificial Swarm Intelligence (ASI) and Conversational Swarm Intelligence (CSI) by demonstrating that real-time, AI-mediated conversational networks can operate effectively in group forecasting domains. Unlike traditional expert elicitation methods that statically aggregate independent judgments, the Hyperchat AI framework enables structured exposure to competing insights that challenge prevailing views, driving reflection and concession, or eliciting counterarguments. The Deliberative Matching Engine plays a central role in this process by optimizing informational diversity and mitigating social influence bias.

While promising, this study had a limited sample size (although sufficient for statistical testing). Future work should be conducted to evaluate performance across a wider range of forecasting domains including financial markets, public health, and enterprise decision-making. Additional research should also explore scaling to larger populations, connecting teams by voice and video conferencing, and enabling hybrid human–AI teams.

In conclusion, the findings support the hypothesis that hyperconnected conversational structures, when mediated by intervening AI agents, can significantly amplify group accuracy in real-world forecasting tasks. By structuring large networked teams into dynamically linked deliberative systems, Hyperchat AI offers a scalable framework for enterprise collaboration across large organizations. This has meaningful implications for organizational decision-making, problem-solving, forecasting, and consensus-building at large scale. In addition, it offers a viable pathway for enabling large, networked human groups to solve challenging problems as a collective superintelligence.